\documentclass[12pt,preprint]{aastex}

\newcommand{\be}{\begin{eqnarray}}
\newcommand{\ee}{\end{eqnarray}}

\begin{document}

\title{Disc-Disc Encounters between Low-Mass Protoplanetary Accretion Discs}

\author{S. Pfalzner}

\affil{I. Physikalisches Institut, Universit\"at zu K\"oln, Z\"ulpicherstr.77,
 50937 K\"oln, Germany}

\author{S. Umbreit, Th. Henning}

\affil{Max-Planck-Institut f\"ur Astronomie, K\"onigstuhl 17, 69117 Heidelberg,
Germany}

\begin{abstract}
Simulations of the collapse and fragmentation of turbulent molecular clouds
and dense young clusters show that encounters between disc-surrounded 
stars are relatively common events which should significantly influence the 
resulting disc structure. In turn this should alter the accretion rate of 
disc matter onto the star and the conditions under which planet formation 
occurs. Although the effects of star-disc encounters have been previously 
investigated, very little is known about encounters where both stars are 
surrounded by discs. In this paper encounters of such disc-disc systems are 
studied quantitatively. It is found that for low-mass discs 
($M_D$= 0.01 $M_\sun$)  the results from star-disc encounters can be
straightforwardly generalized to disc-disc encounters as long as there is 
no mass transport between the discs. Differences to star-disc encounters occur
naturally where significant amounts of matter are transported between the discs.
In this case it is found that although the mass distribution does not change 
significantly, matter caught onto highly eccentric orbits is transported 
surprisingly far inside the disc. The captured mass partly
replenishes the disc, but has a much lower angular
momentum. This can lead to a reduction of the angular momentum in the entire
disc and thus considerably increased accretion shortly after 
the encounter as well as in the long term. 
\end{abstract}


\section{Introduction}
Numerical simulations of the fragmentation of molecular clouds produce
many examples of interactions between fragments with disc-like structures
(Bate et al. 2002) demonstrating that encounters might be
important in the early epochs of star formation. Even in the later
stages of the star formation process, it is found that a star is likely to 
undergo at least one encounter closer than 1000AU during the lifetime of its 
disc (Scally \& Clarke 2001). Recent simulations of the ONC
cluster indicate that at least 20\% of the stars  undergo encounters
closer than 300AU during the first 3Myrs of the cluster development
(Olczak et al. in prep.).\\
Naturally such encounters
will influence the mass and angular momentum distribution in the disc.
The total mass of - and the mass distribution within - the disc after an
encounter are important, as this directly influences the likelihood of the 
formation of planetary systems. The change of the angular momentum 
distribution due to an encounter is of relevance since it is still unclear
how the disc loses enough angular momentum that accretion of 
matter onto the star is possible. Although it is unlikely that
encounters are the dominant source of angular momentum loss, their
contribution is probably not negligible.\\
Because encounters where only one of the stars is surrounded by a disc 
- so called star-disc encounters - are less complex than the situation, where
both stars are surrounded by discs - in the following called disc-disc 
encounters -, it is not surprising that past 
investigations have concentrated on star-disc encounters. 
To our knowledge only Watkins et al.(1998a,b) investigated encounters 
between two disc-surrounded stars. However, they focussed on the likelihood of
the formation of multiple systems in the case of heavy discs and
not on the discs as such.\\
Over the last ten years, the investigations of 
star-disc encounters have progressed from distant encounters 
(Ostriker 1994, Larwood 1997), and a qualitative approach
of closer encounters (Heller 1995, Hall et~al. 1996, Boffin et al. 1998,
Watkins et al. 1998a,b)
to a more quantitative knowledge of how the mass distribution and the angular
momentum in the disc are affected by star-disc encounters
(Pfalzner 2003, 2004).\\
However, as stars tend to form in clusters, it is at least as likely - perhaps
even more likely - that in such an encounter {\it both} stars are surrounded by
a disc. Most of the above mentioned star-disc encounter investigations
can only speculate on the extent to which their results
would be applicable to encounters where both stars are 
surrounded by a disc. \\
It is the aim of this paper to clarify this point quantitatively and to
determine
the conditions for which the results from star-disc encounters can be generalized
to disc-disc encounters.
In this paper, tidal effects in encounters between two stars both
surrounded by low-mass discs are studied numerically in detail. 
A systematic study will be presented for low-mass discs, where
simple N-body simulations suffice.
It is demonstrated that for encounters of low-mass disc without
mass exchange between the discs, there is no real difference to star-disc 
encounters in terms of mass and angular momentum transport. In this case,
the results for star-disc encounters can be generalized to disc-disc encounters. 
Moreover, for the case where mass is exchanged between the discs, it is 
demonstrated that the additional mass alters the mass distribution only slightly. 
However, because the disc gains mass via the captured material the total mass loss
in the disc is less than in an equivalent star-disc encounter. On the other hand, the angular
momentum is still reduced as the captured mass has a much smaller angular 
momentum than that of the original disc material. 
In section 3.3 and 3.4 the effects of viscosity, pressure and self-gravity on the mass and 
angular momentum loss in heavier discs are examined followed by a discussion in section 4.

\section{Initial conditions and numerical methods}
The results reported here are obtained from the disc simulations using 
10 000 pseudoparticles each as tracers of the observed gas. 
These type of simulations are computationally expensive, so
that higher resolution simulations can only be performed for special cases.
Simulations using 1 million particles (see Fig.1) revealed only minor 
differences in the global features of interest, justifying the choice of 
lower particle number for the parameter studies.\\
\subsection{Initial conditions}  
In the simulations, both stars are allowed to move freely and the masses of 
the stars are varied in the different encounters between 
$M_{*}=0.1M_{\sun}$ and $1.5M_{\sun}$, where in most cases one of the 
stars has a mass $M_{*} = 1 M_{\sun}$. The discs extend in all simulations
to 100 AU. As customary for such simulations there exists 
a inner gap from the star to 10\% of the disc size. The gap of 10 AU 
prevents  additional complex calculations 
of direct star/disc interactions and saves computer time.
In addition, particles are removed from the calculation and added to the 
star mass if they come closer than 1AU to the stars. This overestimates 
accretion but avoids the simulation becoming prohibitively slow in order to 
model the close orbit of just one particle correctly.
In the low-mass case, treated in Section 3, we chose $M_D= 0.01 M_{\sun}$ 
as this is a typical mass value of observed discs around young solar-type stars\cite{mannings}.
The density distribution in the disc 
is given by \[
\rho(r,z)=\rho_{0}(r)\,\exp\left(-\frac{z^{2}}{2H(r)^{2}}\right)\,,\]
where $H(r)$ is the local vertical half-thickness of the disc, which
depends on the temperature, and $\rho_{0}(r)$ is the mid-plane density
with $\rho_0(r)\sim 1/r^2$ giving a surface density of $\Sigma\sim 1/r$.
For the $M_{D}$ = 0.01 $M_{\sun}$ case $\rho_0(r) = 1.3 \cdot$ 10$^{-10}$
g/cm$^3$ at $r$=1 AU. \\
The periastra of the encounters are chosen between 100 and 350 AU.
As the  discs extend to $r_{D}$ = 100 AU, this covers the parameter space 
from  penetrating to distant encounters. \\
As described by Heller (1995) the obtained results can be 
generalized for other mass distribution within the disc by applying 
appropriate scaling factors. 

\subsection{Numerical methods}

Including pressure and viscous forces and/or self-gravity 
increases the computational expense. Therefore,  a test particle model is 
used for the limited parameter studies in the first part of the paper. Here the 
disc particles only feel the force of the two stars, but in order to go 
beyond restricted 3-body simulations, the forces of the gas particles
onto the stars are included too.  This means that particles 
close to the center of the disc can influence the movement of 
the star. Due to their low mass, the influence of single particles 
will generally small. However, if the mass distribution near the center 
becomes in any way asymmetric (for
example by captured material) this can influence the movement of the star. As 
Pfalzner(2003) showed the movement of the central star plays a vital role in 
the formation of the second spiral arm in the disc. Thus there is always a 
feedbck between disc and star.

One could question this approach of excluding pressure and 
viscous forces and/or self-gravity, especially 
in disc-disc 
encounters. However, the comparison  with our simulations that include all 
above mentioned effects (see Section 4) show that this simplification is 
justified as long as one is only studying such global features as mass 
transport and restricts oneself to low-mass discs. 
For higher mass discs ($M_{D} \gg 0.01 M_*$) the simple test particle model 
no longer suffices. In section 4 the effects of pressure and viscous forces
as well as self-gravity are considered. 
Selected smoothed particle hydrodynamics (SPH) simulations are
performed to give an qualitative assessment of these effects 
for high-mass discs in general. Details of the numerical method for both kind 
of simulations are described in Pfalzner(2003).\\
\section{Disc-disc interaction}
\begin{figure}
\includegraphics[%
  width=8.5cm,
  keepaspectratio]{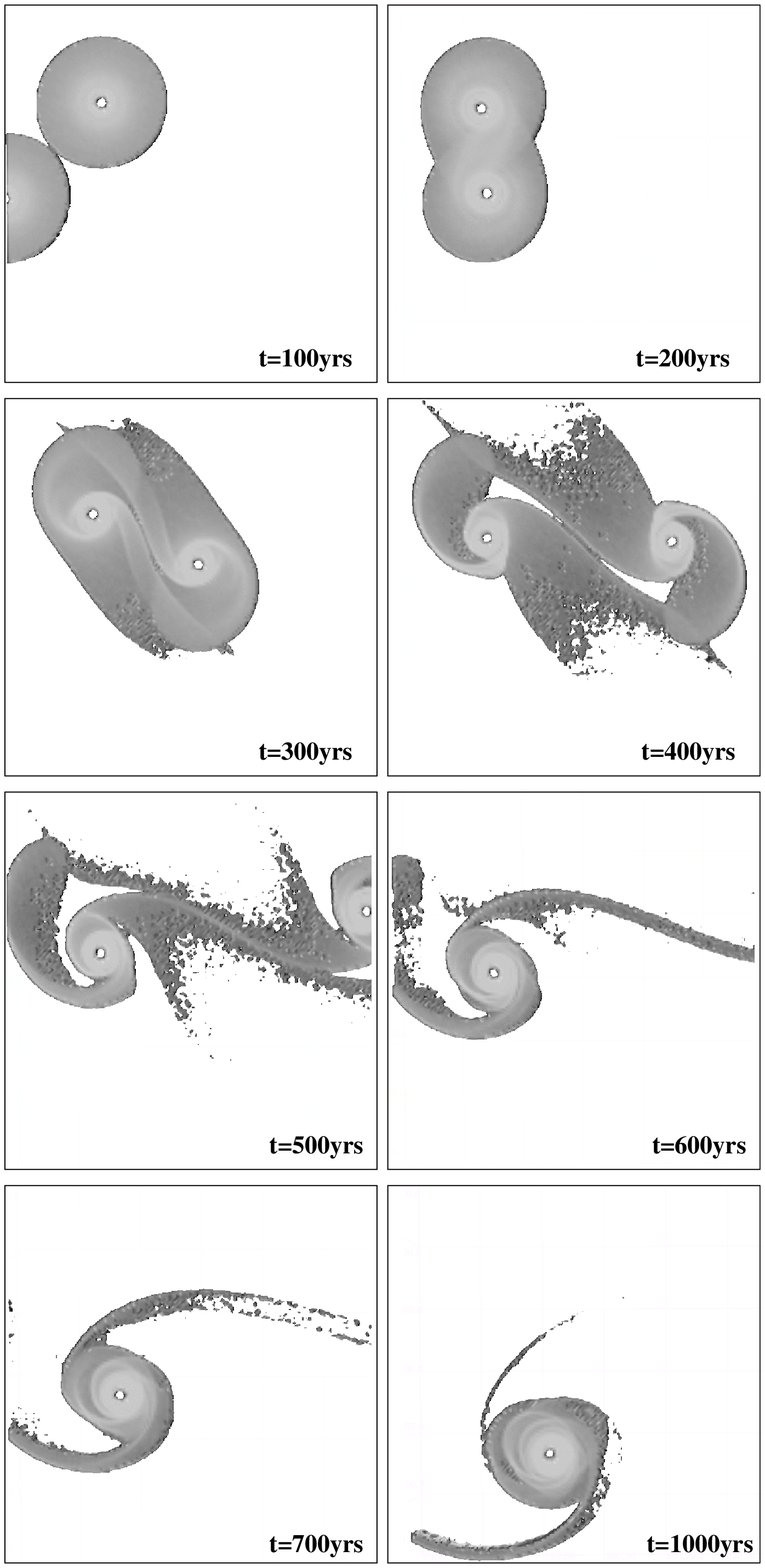}  
\caption{\label{fig:encounter2} Typical example of the simulation of an 
encounter between two low-mass discs. The simulation has been performed
with 500 000 particles per disc. Shown are density plots where the dark areas
represent low density regions.}
\end{figure}

A typical example of an encounter where both stars are surrounded by discs
is shown in Fig.\ref{fig:encounter2}. In this example both 
stars are of equal mass, $M_{1}^*$ = $M_{2}^*$ = 1$M_{\sun}$, as are the 
discs, 
$M_{D}^1=M_D^{2}=0.01M_{\sun}$. The stars move on a parabolic orbit, the
encounter is prograde, coplanar, with both discs rotating in the same 
direction and 
the distance between the stars at the periastron is 150 AU.
After approximately $t$=700yrs a tidal bridge develops, which is equivalent to
the appearance of the first spiral arm in star-disc encounters. The picture at
$t$=800yrs shows the appearance of two counter tidal tails. At 
$t$=1300yrs it can be seen how the disc develops independently towards 
its new equilibrium states. The resulting discs illustrate the 
characteristic tightly wound spiral arm pattern very similar to that seen
in the outer discs recently observed by Grady et al. (2001), 
Clampin et al.(2003) and Fukagawa et al. (2004).\\
We varied the relative periastron distance, $r_{peri}/r_{disc}$, and eccentricity of 
the 
orbit in our simulations of such encounters. The mass and angular momentum 
distribution as well as the total mass and angular momentum loss in the disc after 
the encounter were analyzed and the results of disc-disc compared with
those of star-disc encounters.
\subsection{Mass distribution}
It is important to know the total disc mass as well as the mass distribution 
within the disc after the encounter in the context of planet formation. If 
these quantities are considerably changed by encounters, this could have 
serious implications 
for the possible development of planetary systems. For example, in the
extreme case of a very close encounter the stars might be completely
stripped of their discs  or the disc could become truncated at a certain radius,
thus prohibiting planet formation either altogether or beyond a certain radius. 
In the more likely case of a distant encounter, it is known from star-disc 
encounter simulations that the central region of the disc remains relatively 
unaffected, but the perturbation of the outer disc could lead to planets on 
highly eccentric orbits. The recently discovered planetoid 2003 VB12 
(Trujillo et al. 2004) on an eccentric orbit with a perihelion distance of 70AU 
might possibly have formed through such an interaction (Keynon \& Bromley 2004).

\begin{figure}
\includegraphics[%
  width=8.5cm,
  keepaspectratio]{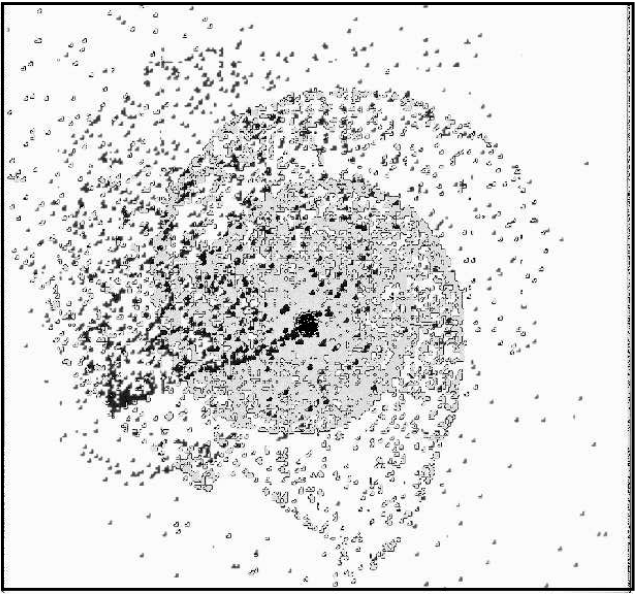}  
\caption{\label{fig:closeup}Closeup of one of the discs at t= 1300yrs
after an parabolic encounter with $M_{1}$ = $M_{2}$ = 1$M_{\sun}$ and
 $r_{peri}/r_{disc}$=1.5. The particles in grey are from the original disc of 
the considered star, whereas the particles in black originate from the disc 
of the other star.}
\end{figure}

\begin{figure}
\includegraphics[%
  width=8.5cm,
  keepaspectratio]{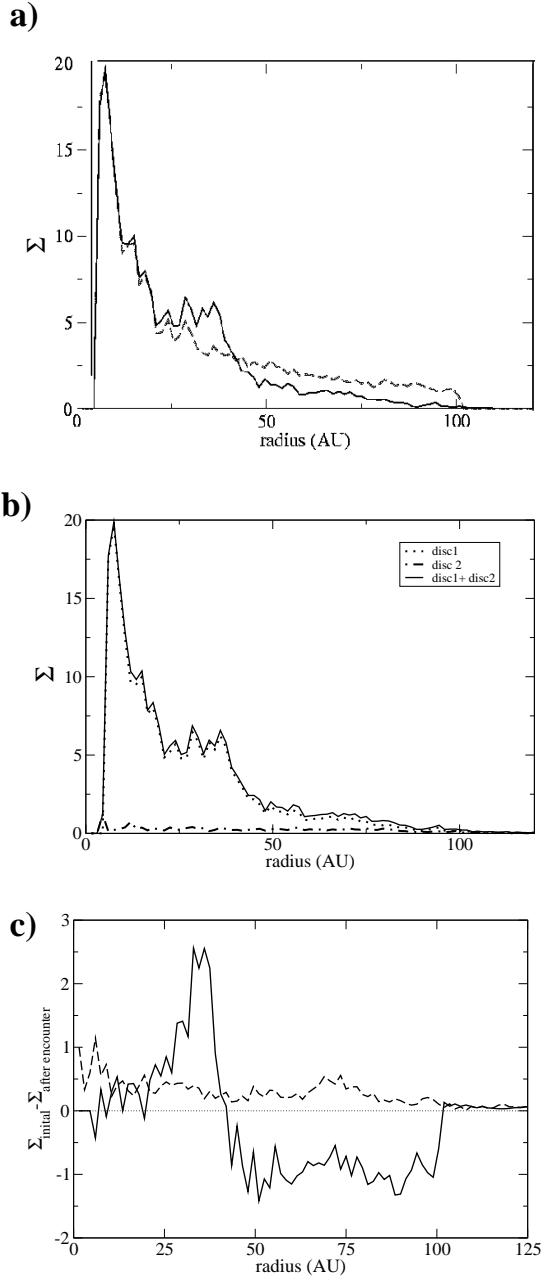}  
\caption{\label{fig:change}Change of the surface density $\Sigma$ in the disc 
as a function of the radial distance. The surface density is given in units of
10$^{-6}$ M$_\sun$/AU$^2$ corresponding to 8.9 g/cm$^2$. In a) the original 
surface density is shown as dashed line, whereas the surface density 
distribution after the encounter is shown as solid line. b) shows how
the total surface density (solid line) after the encounter contains
matter originally belonging to disc 1 (dotted line) and matter originating
from disc 2 (dash-dotted line). It demonstrates that the contribution
from the disc 2 is relatively small. However, in the change
of the distribution ($\Sigma_{\mbox{initial}}$-
$\Sigma_{\mbox{after encounter}}$) the particles from disc 2 (dashed line) 
play a non-negligible role compared to those from disc 1 (solid line) as 
c) shows. }
\end{figure}

Fig.\ref{fig:closeup} shows a closeup of the snapshot at t= 1300yrs of an 
encounter similar to  Fig.\ref{fig:encounter2} but with a periastron distance 
of 150 AU and with the smaller spatial resolution of 10 000 particles
per disc; which illustrates the mass originating from disc 1
(grey) and the mass that star 1 captured from disc 2(black). As the captured 
mass is clearly visible throughout the entire disc, this seems to imply a 
quite altered
mass distribution. In Fig.\ref{fig:change}a the initial mass distribution and 
the distribution 1200 yrs after the encounter are shown as a function of the radial 
distance from the star for the example shown in Fig.2. The comparison of
these two distributions gives quite a different impression than 
Fig.\ref{fig:closeup}. As in star-disc 
encounters, mass is lost from the outside of the disc and also transported 
inwards (for a detailed explanation of the dynamics see 
(Pfalzner 2003). Fig. \ref{fig:change}b shows that the amount of 
mass caught from the other disc is nevertheless relatively small compared to 
the total disc mass, showing that the resulting mass distribution does not differ
significantly from that in star-disc encounters.
Fig.\ref{fig:change}c shows that the gain in surface density through 
this mass from the passing star-disc system is nearly constant at all disc 
radii, even showing a small rise towards the center. Since the above situation 
might not be a typical example, we compared the change in the mass distribution 
in star-disc with that in disc-disc encounters for several different encounter 
situations varying the periastron and the star masses. The result is that the  
mass distributions after the encounter are 
(also quantitatively) like those in star-disc encounters
in all investigated cases.  This is found both for prograde as well as 
retrograde encounters. In cases where there is a change in the density profile 
and/or a truncation of the disc in star-disc encounter, this will equally apply to
disc-disc encounters. Results concerning the mass distribution
from star-disc encounters can therefore be generalized for disc-disc encounters.
Investigations of the mass distribution in star-disc encounters can for example 
be found in Pfalzner(2003); the disc size in such a situation has been
analyzed in Pfalzner et al.(subm).

However, not all is the same for star-disc and disc-disc encounters.
Fig.\ref{fig:change}c compares the mass losses from the disc
with the mass gained from the other disc as a function of the radial distance 
from the star. It can seen that in this example the 
captured mass is only about a factor 3 smaller than the mass transported 
within the disc itself. More importantly, the captured material is partly 
transported very far inside the disc.
This can have two consequences i) the possibility of mixing 
material of different chemical composition and ii) influencing the movement
of the central star and therefore the further development of the entire disc.
Both effects would need further investigation.\\

This raises the obvious question: 
Under which conditions is mass transported between both discs? To answer 
this question we performed a number of encounter simulation with different
encounter parameters.

\begin{figure}
\includegraphics[%
  width=8.5cm,
  keepaspectratio]{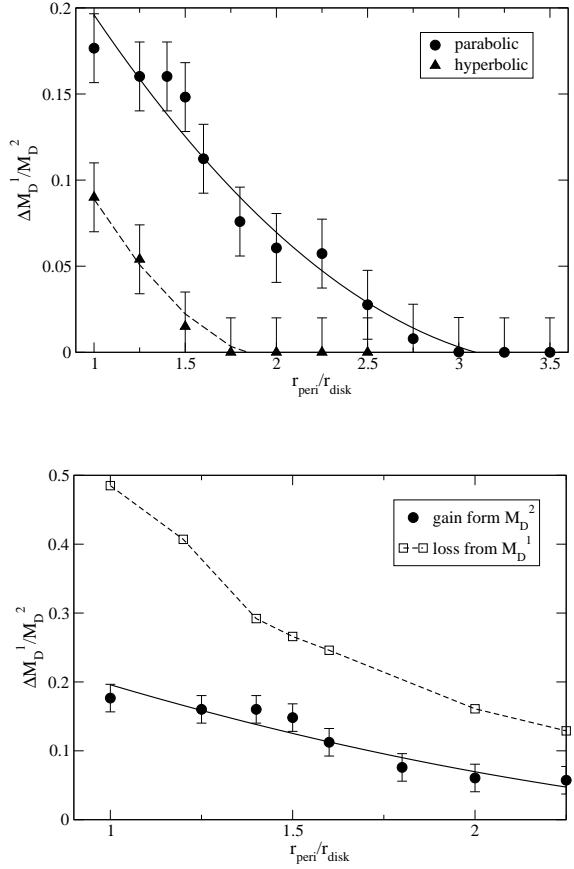} 
\caption{\label{fig:exch} Mass transferred from the disc around star 2 to the
disc around star 1 in an encounter $M_{1}$ = $M_{2}$ = 1$M_{\sun}$ as
a function of $r_{peri}/r_{disc}$. In a) the mass transfer in parabolic 
encounters is compared to that in hyperbolic encounters. 
b) shows a comparison of the loss of matter originally belonging to disc 1
with the gain of matter through capture of matter orignally belonging to disc 
2.}
\end{figure}

In Fig.\ref{fig:exch} the mass of the disc around star 2 captured by star 1 
is shown in encounters for different periastron distances, where both stars 
are of equal mass,$M_{1}^*$ = $M_{2}^*$ = 1$M_{\sun}$.
It can be seen that for hyperbolic encounters, a much 
smaller mass transfer occurs between the disc than for parabolic encounters -
which can be up to 20 \%. This is caused by the fact that the interaction time 
in a hyperbolic 
encounter is much shorter than in an parabolic one, leaving less time for 
the star to actually bind matter. This raises the question, whether
in a parabolic encounter the loss of disc material can be compensated by the 
gain of mass from the other disc.

To answer this we compare the mass loss from disc 1 to the gain in 
matter captured from disc 2. In Fig. \ref{fig:exch}b it can be seen that 
in the parameter range of $r_{peri}/r_{disc}$ = 1 - 2.25 the mass loss is  
at least two-times larger than 
the gain through captured mass: in other words the gain in disc mass  
cannot completely replenish the disc. It can also be seen that the difference 
becomes larger for closer encounters.

\begin{figure}
\includegraphics[%
  width=8.5cm,
  keepaspectratio]{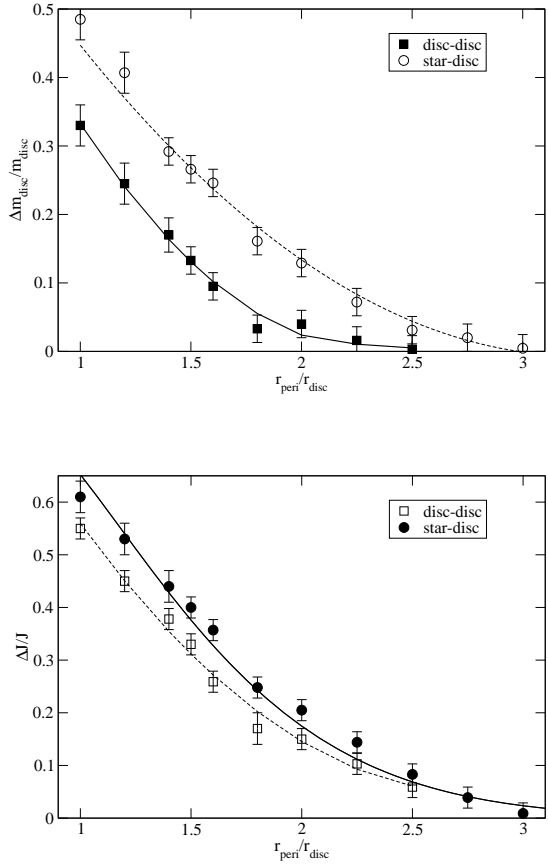}  
\caption{\label{fig:peri} Dependence of the a) mass and b)
angular momentum loss in the disc as a function of  $r_{peri}/r_{disc}$ for 
an encounter of two stars of  $M_{1}$ = $M_{2}$ = 1$M_{\sun}$}.
\end{figure}

In a second step we followed this point up by measuring the mass loss and 
angular momentum loss in star-disc and disc-disc encounters for 9 different 
periastron distances. Fig.\ref{fig:peri} compares the total mass loss in 
star-disc encounters to that in disc-disc encounters. Not surprisingly due to 
the mass gained from disc 2, the mass loss in disc-disc encounters is 
generally less than that in star-disc encounters. 

\begin{figure}
\includegraphics[%
  width=8.5cm,
  keepaspectratio]{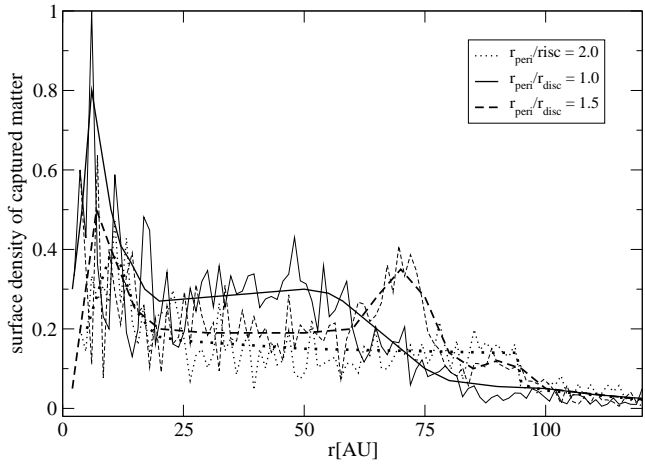}  
\caption{\label{fig:captdist}
Surface density of the captured mass as function of the radial
distance from star 1 after encounters with $M_*^1=M_*^2=1M_\sun$ and 
a periastrion distance of 100AU, 150AU and 200AU.
}
\end{figure}

It is curious that the mass exchange between the disc significantly alters the
total mass loss, but does not seem to affect the mass distribution so much. 
The reason for this is the way the captured mass is distributed.
Fig.\ref{fig:captdist} shows the surface density of the captured mass for 
different encounter periastra. Obviously the captured mass is concentrated 
closer to the star for smaller encounter parameters. This means when more 
mass is captured, most of it is hidden in the steep density gradient 
near the central star, whereas only a 
small amount of the captured matter resides in the outer regions of the disc.

\subsection{Angular momentum}

The angular momentum and the angular momentum per particle are important in 
the context of accretion: the smaller the angular momentum the easier the
matter can be accreted. 

\begin{figure}
\includegraphics[%
  width=8.5cm,
  keepaspectratio]{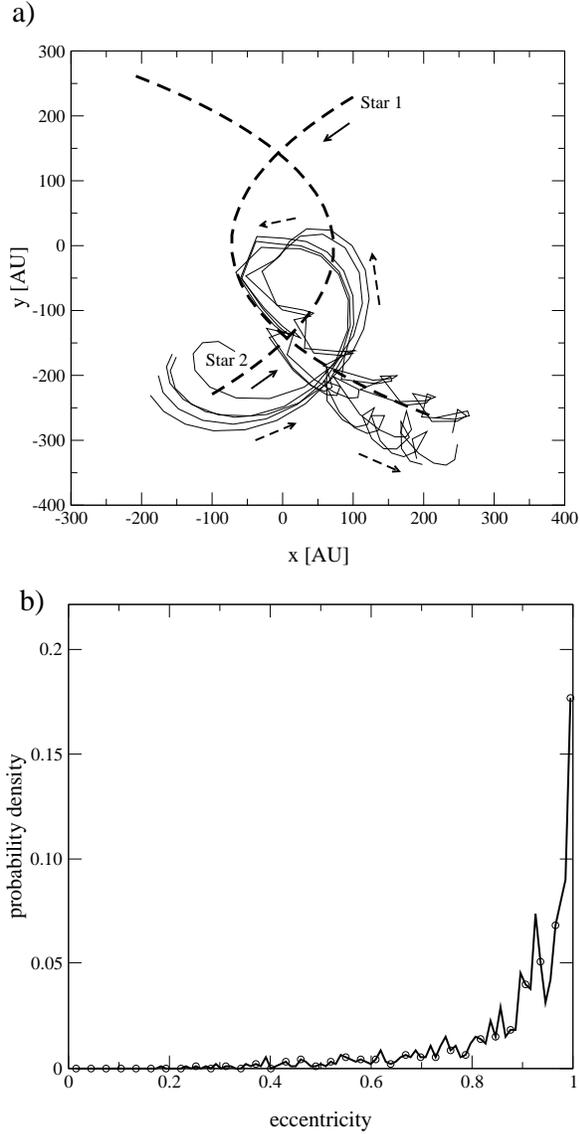}  
\caption{\label{fig:excentricity}a) shows the
trajectories of some particles being captured by the central star (star
1) relative to the center of mass of the stars (solid lines). The
dashed lines represent the orbits of the stars. The captured particles
go on quite complex trajectories and end up in highly eccentric orbits around
star 2 with the same direction of rotation they had before around the
other star.
b) illustrates the eccentricity distribution of the captured particles. Most of 
them end
up in highly eccentric orbits with eccentricities higher than
0.8. Almost none of them has eccentricities below 0.4.
}
\end{figure}

It turns out that for the angular momentum loss in a disc-disc scenario the
trends are considerably different from that of the mass loss. Although the 
angular momentum loss is reduced in comparison to the situation in
star-disc encounters (see Fig.\ref{fig:peri}b), it is so to a much lesser 
degree than the 
mass loss. The reason is that after some complex path in the capturing process
(see Fig.\ref{fig:excentricity}a), the captured particles settle on the above 
mentioned highly eccentric orbits (see Fig.\ref{fig:excentricity}b) around 
the star 1 with the same direction of rotation they had before
around the star 2. Most ($\sim$ 80\%) of the captured particles have 
eccentricities above $0.8$ and almost none of them have orbits less eccentric 
than $0.4$. 

\begin{figure}
\includegraphics[%
  width=8.5cm,
  keepaspectratio]{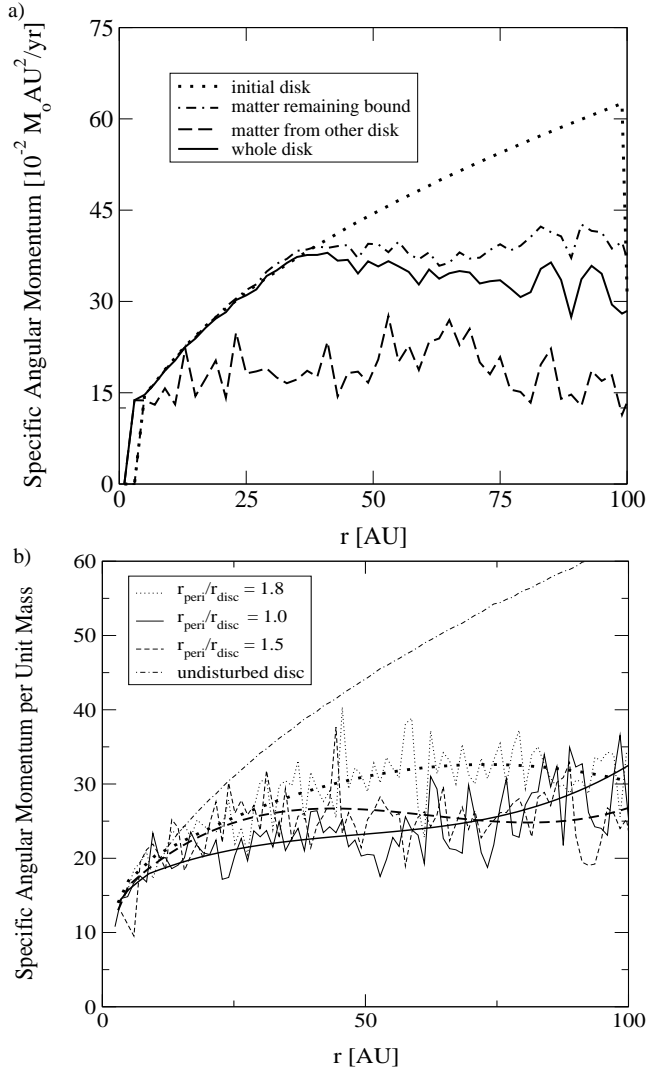}  
\caption{\label{angpart}Change of specific angular momentum per particle
during disc-disc encounters. a) Shows the initial angular momentum per 
particle before the encounter(dotted line), the particles remaining bound to 
the primary (dotted-dashed line), the particles that come from the passing 
star(dashed line) and all particles together(solid line) as a function of the
radial distance from star 1 for the case in Fig.1. b) illustrates the
 specific angular momentum per particle of the captured mass for different
encounter periastra.}
\end{figure}

Consequently the captured particles have a {\it low angular momentum}
relative to the star, 
so that the total angular momentum loss of the disc does not differ much from 
that in star-disc encounters. This becomes even more apparent if one looks at 
the specific angular momentum. In Fig. \ref{angpart}a the specific angular 
momentum per mass unit is shown for the encounter of Fig.2. 
It can be seen that in this particular case, the angular momentum of the 
captured matter is only about half that of the other particles throughout the 
disc apart from the innermost 15 AU. What is particularly striking is 
that  for such close encounters the angular momentum of the particles originally belonging to the
disc is reduced from the disc edge down to about 40AU, whereas the angular momentum 
reduction of the captured material reaches into the disc to about 10-12AU.  
So for very close encounters the determination of the 
angular momentum of the captured material is critical due to the inner cut-off 
of the disc at 10AU. For these cases ($r_{peri}/r_{disc}< 1.1$) simulations
with better resolution of the inner disc area would be necessary.  

Fig. \ref{angpart}b shows the angular momentum per unit mass of the captured 
material in encounters with different periastra. In the outer regions
the angular momentum is always much smaller than in the unperturbed disc 
more or less independent of the encounter periastron. Here 
the angular momentum is always about half that of the unperturbed disc. 
In the inner regions of the disc the reduction seems to become slightly more 
for closer encounters. The simulations probably under-resolve this effect
here, since 
particles that approach the star closer than 1AU are removed from the 
calculation in order to save computer time. These particles might either be 
accreted or go on highly eccentric orbits too.

\subsection{Pressure, viscosity and self-gravity effects}

A common objection to the above approach is the neglect of pressure, 
viscosity and self-gravity in the disc. To justify this simplification
we performed additional simulations including pressure and viscosity effects, 
and then self-gravity.

The initial temperature profile was chosen using the thin-disk 
approximation with a scale height of the disk of $ H(r) = r/H_0=r/$33, 
where $H_0$ is a 
dimensionless constant, and an isothermal equation of state giving $T(r)
\sim$ 1/$r$ . Using the fact that $r/H = v_\phi/ c_s$, with $c_s$ being the 
isothermal sound speed and $v_\phi$ the velocity component of the disk 
material in $\phi$ direction, 
$H_0$ simply represents the local Mach number of the disk material. 
Viscosity was included in the SPH scheme using the tensor 
description by Flebbe et al.(1994). For simplicity a constant kinematic 
viscosity $\nu=4.6\cdot10^{6}\rm{cm^2/s}$ was used, which is related to the 
parameter $\alpha$ in the standard $\alpha$-model by $\nu = \alpha c_s(r) H$
with $c_s$ being the isothermal sound speed. Since $c_s$ is decreasing
as $1/\sqrt{r}$, the same radial dependence holds for $\alpha$ with a
value from $\alpha$=0.16 at 5AU to  0.036 at 100AU. 
The SPH-code itself is described in detail in Pfalzner (2003).

To compare the simulation without and with pressure and viscous effects is 
not straightforward. The problem is that there are three different competing
effects occuring: altered initial distribution, pressure and viscous effects
and star-disc vs. disc-disc encounter, which can influence the final outcome.
The first point arises from the necessity of starting the 
simulation with a disc that is as close as possible to an equilibrium state 
including pressure and viscous forces. This disc usually deviates 
from the Keplerian disc with a 1/$r$-surface density distribution: the higher the 
temperature the more so.
Consequently,  the initial mass and velocity distribution in the encounter 
scenario is in this situation always slightly different to the Keplerian
disc. Although the strongest effects of pressure and viscous 
forces are felt at the center of the disc, the disc edge is also effected:
essentially the outer edge of the disc is more smeared-out and surface 
density near the inner edge is reduced.

It is difficult to distinguish between effects that are due to the 
altered initial distribution with pressure and viscous effects to those that
result from both stars being surrounded by a disc. Therefore we will only 
compare star-disc and disc-disc encounters which start with the same initial 
disc including pressure and viscous forces in the calculation.

\begin{figure}
\includegraphics[%
  width=8.5cm,
  keepaspectratio]{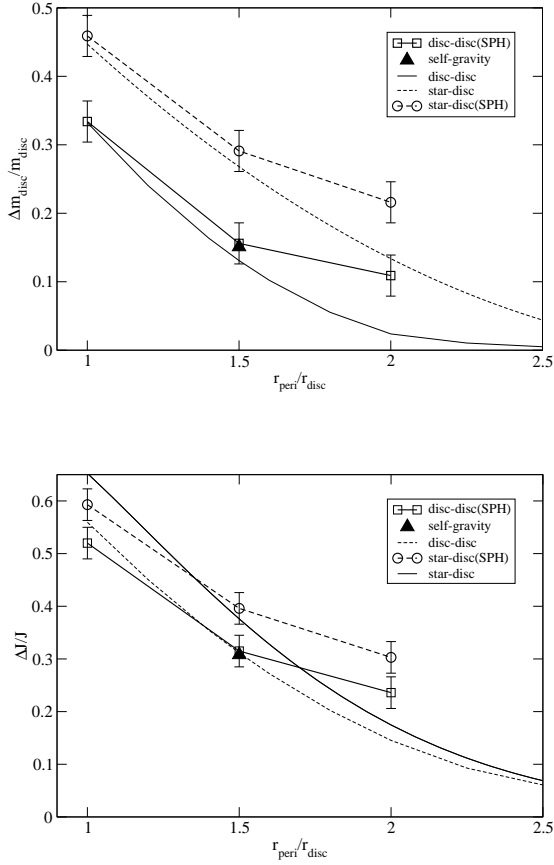}  
\caption{\label{fig:distsph}
Comparison between disc-disc(squares, drawn lines) and star-disc(circles, 
dashed lines) encounters simulations with pressure and  viscous forces included. 
a) shows the mass loss  and b) 
the angular momentum lossas a function of the periastron. In both figures
the fit to the results from the encounters without pressure and viscous forces 
of Fig.5 where marked by drawn and dashed lines without symbols for comparison.
The triangle indicates the result of a simulation where the self-gravity of
the disc is included.}
\end{figure}

Encounter simulations were performed with three different periastra showing
that despite the different initial conditions, the mass 
distribution after the disc-disc encounter differs very little from that of 
star-disc encounters. In Fig.\ref{fig:distsph} the mass and angular momentum 
loss of these gaseous, low-mass discs is shown as function of the encounter 
periastron for both star-disc and disc-disc encounters. For comparison the 
curves fitted to the results of the simulations {\it without} pressure and 
viscous forces (Fig.\ref{fig:peri}) are shown, too. It can be seen that for 
encounters with $r_{peri}/r_{disc}\simeq$ 150 AU, there is little difference 
between the mass and angular momentum loss with and without pressure and 
viscous forces for both the star disc and the disc-disc cases. 

For $r_{peri}/r_{disc}$= 200AU the mass and angular momentum loss are both
considerably higher in the gaseous case. In such distant encounters it is 
mainly the outer regions of the disc which is affected. Since
the mass distribution of the initial disc is more smeared out at the edges, 
there is some matter outside 100 AU. This matter is easily detached from 
the star even in an  200 AU encounter. The reason for the higher mass and 
angular momentum loss in an  200AU encounter for gaseous discs
is just the changed mass and velocity distribution in the initial discs.

Although the periastron dependence of the mass and angular momentum loss
seems to be different for the case of gaseous discs, the relative difference 
between star-disc and disc-disc encounters is to be
the same as in the non-gaseous case. We conclude that viscous and pressure 
forces seem to play a minor role in this context, which is not surprising
given that pressure and viscosity effects predominantly alter the dynamics 
close to the stars. 

Interestingly, pressure forces do not hinder the captured matter from being 
transported far inside the disc. There might be two reasons for this: 
a) the pressure is not high enough to hinder the captured material from 
coming far inside or b) the treatment of viscosity in SPH is not sufficient. 
The latter can only be tested by performing similar simulations with a 
grid-based hydro-code. 

\begin{figure}
\includegraphics[%
  width=8.5cm,
  keepaspectratio]{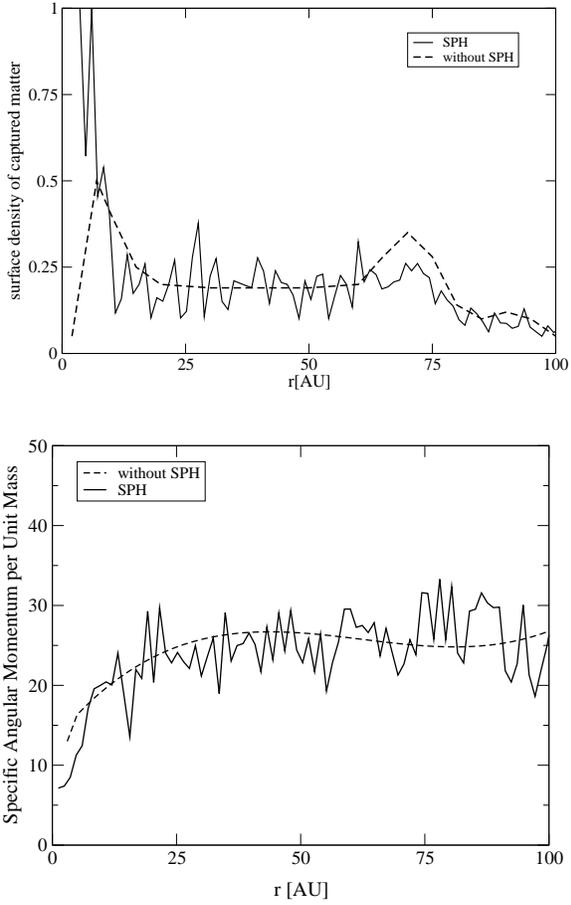}
\caption{\label{fig:distsph2}
Comparison of encounter simulations with (SPH) and without pressure and viscous
forces. In the encounter both stars were surrounded by a disc and the
encounter was parabolic with a periastron of 150AU. a) shows the 
surface mass distribution of the captured material in both cases. b)
shows the distribution of the specific angular momentum in both cases.
}
\end{figure}

Fig. \ref{fig:distsph2}a shows that the captured mass is distributed
in a similar fashion to the case without pressure and viscous forces.
The main difference is that there is no maximum at around 70 AU.
This might be due to the changed initial mass distribution at the outer edge. 
One would expect a 
larger difference in the inner disc where viscous and pressure effects
are strongest. However, there seems to be no indication of that, apart from the
fact that there are slightly fewer particles within 1 AU around the star and 
slightly more in the adjacent areas. The angular momentum distribution also
seems little effected, but this may change on longer time scales than 
the present snapshot 2000 yrs after the encounter. To obtain better information about the long term fate of the matter captured
close to the star, simulations including the inner 10AU area would be
required.

The self-gravity of disc particles is simulated by evaluating the
mutual interactions of all disc particles using a hierarchical tree code
developed by Pfalzner(2003).
Performing a simulation of the same encounter as before, but
including the self-gravity within the disc, our simulations show that 
self-gravity neither influences the mass nor angular momentum transport 
for low mass discs.  The results for the disc with  $M_{D}=0.01 M_{\sun}$ 
including the self-gravity within the disc are virtually identical to those 
of the previous section as indicated in Fig.\ref{fig:distsph}. 

Recently, indications for high-mass discs around more massive stars 
have been found (Shepherd et al. 2001, Chini et al. 2004, Schreyer et al. 2002),
so it is naturally of interest to see what changes to our previous
arguments can be expected for higher masses.
As an example, we choose a disc of mass $M_{D}/M_{*}=0.1$. Again,
an equilibrium solution for the disc has to be found first before 
allowing the two systems to interac. The most relevant effects are the
smeared out edge and a slight increase in velocity in the outer parts of the 
disc (for a comparison of the initial conditions see Pfalzner 2003).

Simulations of such an encounter show that star 2 seems to influence the 
disc of star 1 at a somewhat earlier time. The reason for this lies simply 
in the different initial condition - there is no strict cut-off at 100 AU as 
before, but the edge is smeared out and particles further out are affected 
sooner. In addition, the increased disc mass influences the stellar orbits 
and the higher amount of captured matter contributes to this effect as well.

Interestingly, the time over which mass and angular momentum exchange occur 
after the encounter seems to be longer than for the low-mass case. 
This prolonged exchange time seems to be directly connected
to the larger disc mass - the exchange of energy and angular momentum 
results in a longer period until an equilibrium is reached. 

We find the same behaviour for the angular momentum and mass transfer as in 
above described simulations apart from a much more pronounced 
``eigen-evolution'' of the material remaining bound to the star, which is
typical for self-gravitating discs.  

\section{Discussion and Conclusions}

In this paper the consequences of encounters between two disc-surrounded stars
have been investigated. The simulations results have been compared to the
well-investigated scenario where only one of the
stars is surrounded by a disc. The majority of the results were obtained by
performing simulations without including pressure, viscous or self-gravity 
effects, but additional simulations discussed in section 3.3, which do include
these effects indicate that the main findings still hold
when these effects are included. 

The emphasis of this investigation was
on the mass and angular momentum transport in such disc-disc encounters.
It was found that for low-mass discs, the findings of star-disc encounters 
can be generalized as long as there is no significant mass exchange between 
the disc. 
It has been shown for coplanar encounters with $r_{peri} >$ 100 AU that 
mass exchange between discs can only be expected for nearly parabolic, 
close prograde encounters. For hyperbolic encounters with a periastron larger 
than 3 times the disc size, mass transport between the discs will be 
negligible. This means for the majority of hyperbolic encounters the results 
for the mass and angular momentum transport of star-disc encounters can be
applied to disc-disc encounters.

In cases where matter is transported (mainly parabolic encounters), the 
additional mass alters the mass distribution only slightly. This has several 
reasons: For distant encounters the mass exchange is at most a few percent. 
This is spread more or less evenly over the whole disc area, thus having 
little effect on the overall mass distribution. For close encounters we find 
that up to 20 per cent of disc mass is captured. However, a considerable 
amount of this ends up very close to the star thus ``hiding'' in the steep 
density gradient near the star. On the other hand, because the disc gains 
mass via the captured material, the total mass loss
in the disc is actually less than in an equivalent star-disc encounter. 

The change in angular momentum is more subtle, since affected differently as 
the captured mass moves on highly eccentric orbits, getting transported far 
inside the disc and so  has a much smaller angular momentum than the original 
disc material.  Unlike the matter in the initial disc the angular momentum of 
the captured matter does not increase on average with larger distance from 
the star, but is nearly constant at all radii. 

What consequences do these results have for the late stages of star formation 
and planet formation?

In the context of star formation the accretion rate onto the star is of
vital importance. Since the angular momentum of the captured matter in 
disc-disc encounters is so much lower than that of the initial disc, 
accretion could be enhanced in two ways: 
First, as a considerable amount of the captured matter goes on orbits of 
high eccentricity very close to the star, this can lead to an increase of 
accretion shortly after the encounter.
Second, long term viscous processes could lead  to an exchange of 
angular momentum between the captured and original particles in the disc. 
The circularization of the orbits of the captured matter could in this way 
cause a reduction of the angular momentum of all the particles in the disc. 
This lower angular momentum could eventually facilitate the accretion process.

Of course, it would be difficult to actually observe the increased accretion, 
since even a parabolic encounter lasts no longer than a few hundred years. 
What might be observable, however, is chemical mixing, assuming
the two discs are of different chemical composition.  

Here only hyperbolic and parabolic encounters were investigated, but
the results give some indications for cases where bound binaries are formed 
by the star-disc or disc-disc encounter, too: The case where both stars are 
surrounded by a disc would lead to a much larger decrease in angular 
momentum and  this should lead to much higher accretion rates. So if
binaries are formed in such encounters, then they will do so preferrably
in disc-disc encounters instead of star-disc encounters.

The notion that the development of spiral
arms in gravitationally unstable discs might be an important ingredient
in the formation of planets is an interesting possibility (Boss). 
Such instabilities facilitate the formation of planets due to both
the increased density and through a velocity dispersion that
facilitates dust and planetesimal growth (Rice  et al. 2004). The
problem with this hypothesisis that the timescales for low-mass discs to 
become gravitationally unstable are relatively long. By contrast, 
encounters seem to be much more common in dense young clusters than 
previously thought and lead to very similar but much stronger spiral arms on 
very short time scales, and are therefore a much more plausible catalyst for 
planet formation than instabilities alone. An addition advantage is that
in disc encounters spiral arms - and possibly as well the necessary
conditions for the formation of planets - occur even in low-mass discs. 

As was shown in this paper the captured matter can not be expected to  
increase the density in the spiral arms significantly. However, because it 
moves on highly eccentric orbits far inside the disc, the velocity dispersion
in the disc should be considerably increased. So if one accepts that spiral 
arms are an important ingredient for planet formation, then the two-disc 
interaction will be even more efficient than a star-disc encounter in 
creating the necessary conditions.

\bibliographystyle{apj}

%
\clearpage

\clearpage
\clearpage

\clearpage

\clearpage
\clearpage
%
%

\clearpage

\clearpage

\clearpage
%
%

\clearpage

\clearpage

\end{document}